\documentclass{article}

\usepackage{arxiv}
\usepackage{cite}
\usepackage{amsmath,amssymb,amsfonts}
\usepackage{algorithmic}
\usepackage{graphicx}
\usepackage{textcomp}
\usepackage{xcolor}
\def\BibTeX{{\rm B\kern-.05em{\sc i\kern-.025em b}\kern-.08em
    T\kern-.1667em\lower.7ex\hbox{E}\kern-.125emX}}
\begin{document}

\title{NAViDAd: A No-Reference Audio-Visual Quality Metric Based on a Deep Autoencoder\\
%{\footnotesize \textsuperscript{*}Note: Sub-titles are not captured in Xplore and
%should not be used}
\thanks{This publication has emanated from research supported in part by the Conselho Nacional de Desenvolvimento Cientfico e Tecnol\'ogico (CNPq), the Coordena\c{c}\~{a}o de Aperfei\c{c}oamento de Pessoal de N\'ivel Superior (CAPES), the Funda\c c\~ao de Apoio \`a Pesquisa do Distrito Federal (FAPDF), the University of Bras\'ilia (UnB), the research grant from Science Foundation Ireland (SFI) and the European Regional Development Fund under Grant Number 13/RC/2077 and Grant Number SFI/12/RC/2289.}
}

\author{
  Helard Martinez \\
  %School of Computer Science\\
  University College Dublin\\
  Dublin, Ireland \\
  \texttt{helard.becerra@ucd.ie} \\
  %% examples of more authors
   \And
 Andrew Hines \\
  %School of Computer Science\\
  University College Dublin\\
  Dublin, Ireland \\
  \texttt{andrew.hines@ucd.ie} \\
     \And
 Myl\`ene C.Q. Farias \\
  %School of Computer Science\\
  University of Brasilia\\
  Brasilia, Brazil \\
  \texttt{mylene@ieee.org} \\
}

%\author{\IEEEauthorblockN{Helard Martinez and Myl\`ene C.Q. Farias}
%\IEEEauthorblockA{\textit{University of Brasilia} \\
%Brasilia, Brazil \\
%helardb@unb.br, mylene@ieee.org}
%\and
%\IEEEauthorblockN{Andrew Hines}
%\IEEEauthorblockA{\textit{University College Dublin} \\
%Dublin, Ireland \\
%andrew.hines@ucd.ie}}

\maketitle

\begin{abstract}
%Although the research in audio and video quality assessment
%(tackled as individual modalities) is fairly mature, with  several audio and video quality metrics  proposed
%in the last years, there are still several issues to be solved in the area of audio-visual quality. 
The development of models for quality prediction of both audio and video signals is a fairly mature field. But, although several multimodal models have been proposed, the area of audio-visual quality prediction is still an emerging area. In fact, despite the reasonable performance obtained by combination and parametric metrics, currently there is no reliable pixel-based  audio-visual quality  metric. The approach presented in this work is based on the assumption that autoencoders, fed with descriptive audio and video features, might produce a set of features that is able to describe the complex audio and video interactions. Based on this hypothesis, we propose a No-Reference Audio-Visual Quality Metric Based on a Deep Autoencoder (NAViDAd). The model visual features are natural scene statistics (NSS) and spatial-temporal measures of the video component. Meanwhile, the audio features are obtained by computing the spectrogram representation of the audio component.  The model is formed by a 2-layer framework that includes  a deep autoencoder layer and a classification layer. These two layers are stacked and trained  to build the deep neural network model. The model is trained and tested using a large set of stimuli, containing representative audio and video artifacts. The model performed well when tested against the UnB-AV and the LiveNetflix-II databases. %Results shows that this type of approach produces quality scores that are highly correlated to subjective quality scores.
\end{abstract}

%\begin{IEEEkeywords}
%audio-visual, quality metrics, no-reference, distortions, autoencoder, NAViDAd \end{IEEEkeywords}

\section{Introduction}\label{sec:intro}
The great progress achieved by communications technology in the last twenty years is 
reflected by the amount of multimedia services currently available, such as digital television, IP-based video transmission, and mobile services. Among the most popular multimedia services
are IP-based transmission, including video conference (Skype, Google Hangout, Facebook Video,
FaceTime) and on-demand streaming media (Netflix, iTunes, Hulu, Amazon). Yet, it is understood that the success of these kind of services relies on their trustworthiness and the provided quality of experience~\cite{Korhonen2010}. Therefore, the development of efficient real-time monitoring quality  tools, which can quantify the audio-visual experience, is key to the success of any multimedia service or application.  Although the research in audio and video quality assessment (tackled as individual modalities) is fairly mature~\cite{akhtar2017audio}, there are still several issues to be solved in the area of audio-visual quality.

% For instance, modeling how humans perceive audio and video signals is a challenging task, specially when we consider the interaction in the perception of audio and video signals. 
 
 Audio and visual descriptive features have been  used in several applications, such as speech intelligibility and pattern recognition \cite{hines2012predicting, borji2014human}. Their performance naturally relies on how good these features are able to describe the signal characteristics, specially in terms of human perception. In the quality assessment area, there are several audio and video quality metrics that achieve very good performances using audio and video features, respectively,  to predict the perceived quality~  \cite{akhtar2017audio}. 
%It is worth pointing out that the ML mapping \textcolor{blue}{algorithm} is also an important stage of the quality assessment method. 
But, currently there is no feature-based quality metric that estimates the quality of an audio-visual signal, taking into consideration the characteristics the audio and visual components  and their important interactions.

Considering these issues, Machine Learning (ML) paradigms arise as an appealing option to tackle the audio-visual quality assessment problem from a different perspective. Quality assessment methods based on ML are capable of mimicking human reactions to media distortions, instead of explicitly modelling it. Soni \textit{et al.} used a deep autoencoder strategy to design a non-intrusive speech quality assessment method~\cite{soni2016novel}. The proposed metric adopts a two-layer approach to treat speech background noise distortions, using audio features in the form of spectrograms. In the first layer, a speech spectrogram is passed by a two-layer autoencoder to produce a low-dimensional set of new features. A mapping function between the features and subjective scores is found using an artificial neural network (ANN). Results showed that this deep autoencoder approach produced better descriptive features than Filterbank Energies (FBEs) and more accurate speech quality predictions. We believe that deep autoencoders techniques can be used for the particular task of finding ways to describe the audio and video components interact.

%Traditionally, methods that are based on ML are composed of two basic stages: (1) the computation of features describing the media distortion and (2) a mapping of those features into quality scores. As a result, the model learns the complex non-linear function that maps features into quality scores. Nevertheless, to successfully model these complex functions,  we need to design an appropriate feature set, which  is capable of completely describing the audio-visual signal,  and an adequate ML technique that implements the quality mapping function.

\begin{figure}[t!]
\centering
\includegraphics[scale=0.35]{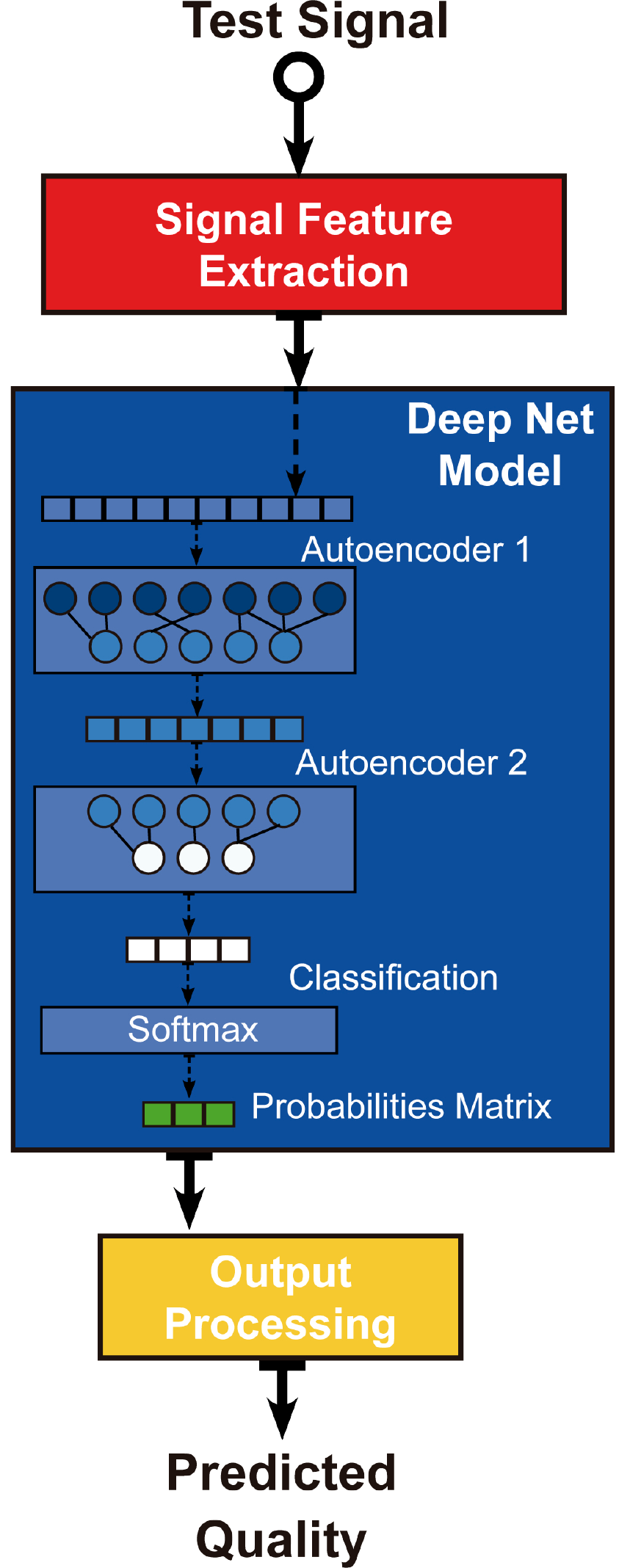}
\caption{Block diagram of the Audio-VIsual Quality metric based on a Deep AutoencoDer (NAViDAd). }\label{fig:block_diagram}
\end{figure}

In this work, we design a No-reference Audio-VIsual Quality metric based on a Deep AutoencoDer (NAViDAd). The proposed model seeks to  blindly estimate audio-visual quality for streaming multimedia applications. To estimate the audio-visual quality, the model uses both audio and video descriptive features. This way, it takes into account not only for the quality of the individual components, but also their interactions. First, a set of features that describe the characteristics of the audio and video components are computed. Next, the set of features are passed to a trained model that is composed of two layers: an autoencoder layer and a classification layer. The autoencoder layer produces a low-dimensional set of features. At this stage, it is expected that these low-dimensional set of features are able to describe the complex interaction between audio and video stimulus. The classification layer is responsible of mapping the features into audio-visual quality scores. Finally, the last stage takes the model output to processed it and calculate the overall audio-visual quality. Figure \ref{fig:block_diagram} depicts the block diagram of NAViDAd, depicting all three stages of the proposed metric.

%Figure
%\vspace{-0.3cm}

The structure of this paper is organized as follows. In Section \ref{sec:arquitecture}, the architecture of the proposed metric is described. In this Section we detail the extraction of the audio and video features as well as the proposed audio-visual quality assessment model. In Section \ref{sec:performance} we present the results obtained with the proposed model. Finally, in Section \ref{sec:conclusion} some conclusions and final comments are presented.
%we detail the extraction of the audio and video features. 

\section{Proposed Architecture}\label{sec:arquitecture}

In this section, we describe the architecture of the proposed NR audio-visual quality metric, which includes: feature extraction,
model training and testing, and output processing. 

\subsection{Extraction of Audio-Visual Features}\label{sec:features}

%the set of video features is a set of natural scene statistics (NSS) used by several image and video quality metrics \cite{zhang2014c, mittal2016completely}, plus the spatial and temporal information associated to each video sequence. As for the audio features, a spectrogram (2-D representation) is used as the audio feature. 
%Next, we describe in more detail the several stages of the audio-visual feature extraction procedure. 

%\subsection{Visual Features}
%As mentioned earlier, the  natural scene statistics and the spatial and temporal information were used as video features, which were used as input for the video quality model. 
%In this work, we extract the visual and audio features independently and, then, merge them to generate the audio-visual features. 
Natural Scene Statistics (NSS) and the spatial and temporal information were used as visual features. We used the feature extract function from the Diivine image quality metric to extract a total of eighty-eight (88) features~\cite{zhang2014c}, resulting in an 88-by-$n$ matrix ($n$ is the number of video frames) that represents the NSS features. To capture the spatial and temporal characteristics of the video, we used the algorithm proposed by Ostaszewska and Kloda \cite{ostaszewska2007quantifying} to compute the spatial and temporal information, helping characterize  important visual distortions (e.g. freezing and packet loss distortions). Spatial and temporal values are computed for each video frame, resulting in a 2-by-$n$ matrix that represents the spatial and temporal features. Both sets of features are merged to form the \textit{visual} features of a video sequence, represented by a 90-by-$n$ matrix. %Figure \ref{fig:visual_feats} depicts the visual set of features for a single video sequence.

%\subsection{Audio Features}
ViSQOL \cite{hines2015visqol} and ViSQOLAudio \cite{sloan2017objective} are full reference speech and audio quality metrics based on the same underlying platform. They use an intensity spectrogram representation of the audio signal, i.e. a time-frequency intensity representation of the audio activity, as the audio feature source for quality prediction. In this work, we use the feature extraction functionality of ViSQOL to process the audio signal and obtain a 25-by-$m$ matrix, where 25 is the number of frequency bands and $m$ is the number of audio frames in the signal.
%Visqol uses the audio signal spectrogram representation, which is basically a time-frequency intensity representation of the audio activity, as the audio feature source. 
%In this work, we used the Visqol feature extraction function, obtaining a 25-by-$m$ matrix, where 25 represents the number of frequency bands and $m$ is the number of audio samples of the signal.
 %Each column of the spectrogram provides a set of 25 descriptive values corresponding to each sample of the audio signal. %Figure \ref{fig:spectr_feats} depicts a sample of the spectrogram matrix extracted from the audio signal.

\subsection{Combination of Audio-Visual Features}
Once the visual features (90-by-$n$ matrix)  and the audio features (25-by-$m$ matrix) are obtained, they are merged together resulting in a total of 115 descriptive features. However, given that the number of video frames ($n$) and the number of audio samples ($m$) are not necessarily the same, a scaling process is required to match these two sets before merging them. To uniformize the length of the two matrices, we simply replicated the values of the matrix that has the shorter length, so that it matches the size of the larger matrix. Since the number of frame videos ($n$) is generally smaller than the number of audio samples ($m$), values of the visual feature set are replicated to match the audio feature set. Once the length of both sets matches, they are merged to form a 115-by-$m$ matrix, denoted as the audiovisual feature set (115 is the sum of the 90 visual features plus the 25 audio features).
%Figure \ref{fig:scale_mats} presents an illustration of this scaling procedure.

%%Figure
%\begin{figure}[t!]
%\centering
%\includegraphics[scale=0.33]{./img/Fig_scale_matrix.pdf}
%\caption{Simplified illustration presenting the scaling procedure to match the visual and audio feature matrices.}\label{fig:scale_mats}
%\end{figure}
%%\vspace{-0.3cm}

Additionally, a \textit{target set} is built using the subjective scores associated with each video under analysis. This set contains the target quality scores used during the model training. Since in an ACR quality scale there are 4 quality groups, which represent the quality intervals assigned to the stimuli, the target set is a 4-by-$m$ matrix composed of zeros and ones, where 4 is the number of quality groups and $m$ is the length of the features matrix (115-by-$m$). The target set is built by taking the average subjective score associated with the stimuli and setting to one the corresponding quality group. For example, a sequence that has a subjective score of 1.65 is assigned the quality group 1 since the score is in the interval $[1,2)$, while a sequence with a subjective score of 3.52 is assigned to the quality group 3 since the score is in the interval $[3,4)$. In summary, the row corresponding to the corresponding quality group is set to one and the rest of the rows are set to zero. Considering that each column represents a video sample, this setup guarantees that each sample has only one quality group associated to it. During the model training, this target set is used to map the corresponding quality group of each sample.
%Figure \ref{fig:target_mat} depicts examples of this setup.

%%Figure
%\begin{figure}[b!]
%\centering
%\includegraphics[scale=0.28]{./img/Fig_target_matrix_intervals.pdf}\\
%\small{(a)}\\
%\includegraphics[scale=0.28]{./img/Fig_target_matrix_examples.pdf}\\
%\small{(b)}\\
%\caption{\textbf{(a)} Target Quality Group Matrix representing the 4 quality group intervals. \textbf{(b)} Sequence with subjective score of 1.65 is assigned the quality group 1, interval [1,2]. Sequence with subjective score of 3.52 is assigned the quality group 3, interval [3,4].}
%\label{fig:target_mat}
%\end{figure}

Then, the feature and target matrices for all the audiovisual signals of the dataset are concatenated to produce two large global sets. The number of rows of the global feature and target matrices are 115 and 4, respectively. Meanwhile, the columns of the matrices are denoted by $M$, which represents the sum of all number of video samples from the training dataset. These two global sets served as input for the model training at different stages. 

\subsection{NAViDAd Training}\label{sec:method}
\begin{figure}[t!]
\centering
\includegraphics[scale=0.35]{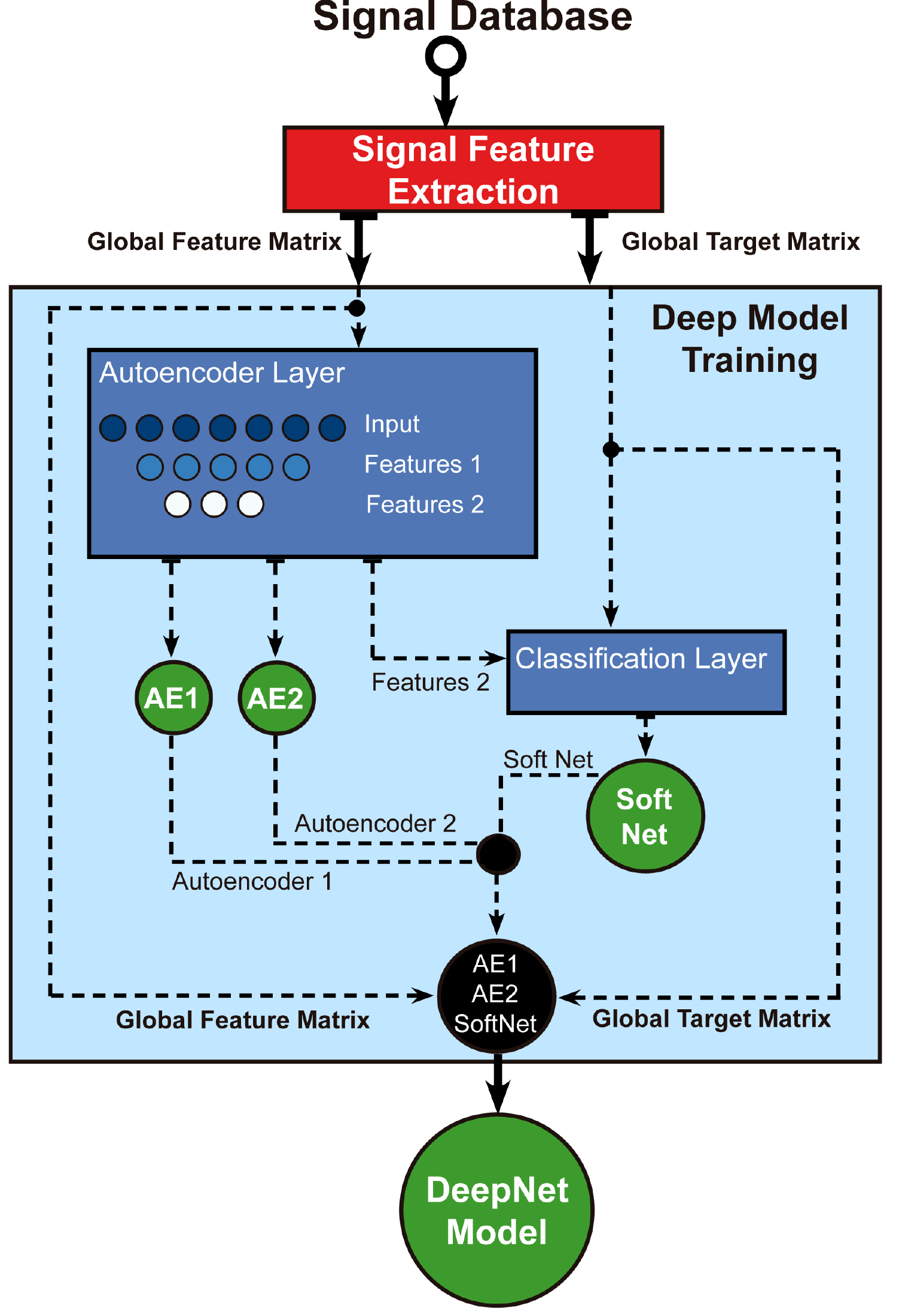}
\caption{Simplified block diagram of the No-Reference Audio-Visual Quality Metric Based on a Deep Autoencoder (NAViDAd).\label{fig:diagramAEModel}}
\end{figure}

A basic block diagram of the proposed No-Reference Audio-Visual Quality Metric Based on a Deep Autoencoder (NAViDAd) is presented in Figure \ref{fig:diagramAEModel}. The training phase consists of two main layers: 1) the autoencoder layer, that receives as input the global feature set, and 2) the classification layer, which receives the global target set and a low-dimensional set of features. Once trained, the resulting models are stacked and trained to build the final deep audiovisual quality model. 

\subsubsection{Autoencoder Layer}
The autoencoder layer produces a low-dimensional set of features that are able to describe the audio and video characteristics, as well as the distortions associated with the signal. Two sub-layers (autoencoders) are used in this phase. {It is worth mentioning that this is a demonstrative structure, further tests might add more sub-layers depending on the requirements of the model.} The first layer receives the global 115-by-$m$ matrix containing the audiovisual features from the training stimuli. This first auteoncoder is trained using a hidden layer of size 60, generating as output a 60-by-$m$ matrix denoted as \textit{Features 1}. Another output of this layer is a trained autoencoder, denoted as \textit{Autoencoder 1}. The second autoencoder has a hidden layer of size 25 and is trained using as input the \textit{Features 1}. It generates as output a 25-by-$m$ matrix,  denoted as \textit{Features 2}, and a trained autoencoder, denoted as \textit{Autoencoder 2}. The training parameters for the  two autoencoders are the following: a linear transfer function is set for the decoder, the L2 weight regularizer is 0.001, the sparsity regularizer is  4, and the sparsity proportion is 0.05. 
%Table \ref{tab:autoenc-params} depicts some additional parameters considered for the training of the model.

In summary, the autoencoder produces: two trained autoencoders (\textit{Autoencoder 1} and \textit{Autoencoder 2}) and two sets of features (\textit{Features 1} and \textit{Features 2}). From these outputs, only \textit{Features 2} is used as input to the following layer: the classification layer.  \textit{Autoencoder 1} and \textit{Autoencoder 2} are used during the overall training of the deep neural network model.

\subsubsection{Classification Layer}
This layer has the goal of finding a mapping function between the input feature set of the training stimuli and the  corresponding subjective scores. To obtain this mapping, a softmax function is used to discover the quality group corresponding to the set of features. This layer receives the \textit{Features 2} set  obtained in the previous layer and a 4-by-$m$ target set and performs the training of the classification function. The resulting function, denoted as Soft Net, is trained to generate a matrix containing the probabilities that a certain sample belongs to each quality group.
After the autoencoders (\textit{Autoencoder 1} and \textit{Autoencoder 2}) and the classification function (Soft Net) are trained, they are stacked to form the deep neural network (Deep Autoencoder Network). Finally, the network is trained using the global feature and the global target sets gathered in the previous layers. 

{A more detailed description of the entire training procedure and information regarding the parameters used for training the model can be found at \cite{phdthesisHelard2019}.}

%%Figure
%\begin{figure}[bt]
%\centering
%\includegraphics[scale=0.33]{./img/Fig_processing_output.pdf}
%\caption{Simplified illustration of the output processing stage applied to the results of the \textbf{Deep Autoencoder Network} model.}\label{fig:output_proc}
%\end{figure}

\subsection{NAViDAd Testing}\label{sec:perform}
To extract the audio-visual features of the test stimuli, the testing stage uses the same procedure used in the training stage. The  extracted set of features (again, a 115-by-$m$ matrix) are, then, passed to the deep autoencoder network. Next, the output (4-by-$m$ matrix) is processed to compute the audiovisual predicted quality. First, the maximum value of each column and its corresponding row index in the 4-by-$m$ matrix are computed. Then, a 1-by-$m$ vector is built by adding the index  and the maximum  value, i.e. for each column the corresponding quality group index is summed to the corresponding probability value resulting in a quality (real) number in the interval [1, 5]. Finally, the quality scores of all $m$ columns are averaged and the overall audiovisual quality score is computed. 
%, which is illustrated in Figure \ref{fig:output_proc}

\section{Performance Analysis}\label{sec:performance}

The NAViDAd model was trained and tested using sequences from the UnB-AVQ database. The UnB-AVQ is a large dataset of audio-visual stimuli (video sequences with accompanying audio) with their corresponding quality scores \cite{martinez2019analyzing}. In this work, we used the third part (Experiment 3) of this dataset, which contains a total of 800 sequences with combinations of audio and video distortions~\cite{martinez2019analyzing}. The video distortions were Bitrate compression, Packet-Loss, and Frame-Freezing. The audio distortions were: Background noise, Chop, Clip, and Echo.  Detailed information regarding the experiment procedure and the distortion parameters can be found in our previous work~\cite{martinez2019analyzing}.
%The Background noise contained  `car' and `office' noises, which were  to the  original signal at two different SNR levels (15 and 10 dB). Clipping was generated by amplifying the signal using a multiplying factor (11 or 25). Echo  was produced by adding  delayed versions of samples (by 100 and 180 ms) to the original signal.  Chop was generated  by substituting the missing samples by zeros, with a period of 0.02s and a frequency of 2 or 5 chops/s. 

We used a 10-fold cross-validation method to train and test the proposed metric.  We compared NAViDAd with the following quality metrics: 
\begin{itemize}
\item FR visual metrics: SSIM and PSNR;
\item NR visual metrics: DIIVINE, VIIDEO, BIQI, NIQE, and BRISQUE;
\item FR audio metrics: VISQOLAudio, PEAQ, and VISQOL (speech);
\item NR Audio metric:  P.563 (speech);
\item NR Audio-visual metrics: Linear, Minkowski, and Power  models, using DIIVINE and P.563.  
\end{itemize}
It is worth pointing out that these metrics were designed for a variety of contexts and were trained with different content. With respect to the audio-visual metrics, given the limitation of space, we have chosen to combine the outputs of the best performing NR audio and video quality metrics.

Table \ref{tab:av_video_results} presents the Pearson and Spearman correlation coefficients (PCC and SCC, respectively), along with the root mean square error (RMSE) obtained for the considered visual and audio-visual quality metrics. Results are organized according to the type of video distortion, given that the visual quality metrics cannot differentiate audio degradations in the stimuli. From Table \ref{tab:av_video_results}, it can be observed that NAViDAd has the best overall accuracy performance (All), when compared to the other visual quality metrics. As for the audio-visual combination models, NAViDAd also shows a clear advantage. When we consider the type of visual distortion, NAViDAd presents the best performance for both frame freezing (0.91) and packetloss (0.86) distortions.  Notice that several visual  (SSIM, NIQE, BRISQE) and audio-visual quality metrics present a lower performance for one type of visual distortion, while NAViDAd achieves a consistent performance.  %\textbf{ACHO QUE SERIA INTERESSANTE COLOCAR as Metricas audio-visuais aqui! }

Table \ref{tab:av_audio_results} presents the results for audio and audio-visual metrics. This time, the results are separated by the audio distortions, given that the audio metrics cannot differentiate video distortions in the stimuli. From Table \ref{tab:av_audio_results}, it can be observed that NAViDAd has the best overall accuracy performance (All), when compared to the  audio and speech quality metrics. This advantage was expected since audio and speech metrics use only the audio component  to predict the perceived quality. As for the audio-visual combination models, NAViDAd also shows a clear advantage.  Regarding the audio distortions, interestingly, NAViDAd presents a better performance for chop and echo distortions (0.92 and 0.90). With regard to the audio-visual combination models, it is clear that NAViDAd performs better and shows a clear advantage.

\begin{figure*}[bt]
\center
\begin{tabular}{cccc}
\includegraphics[width=0.23\columnwidth]{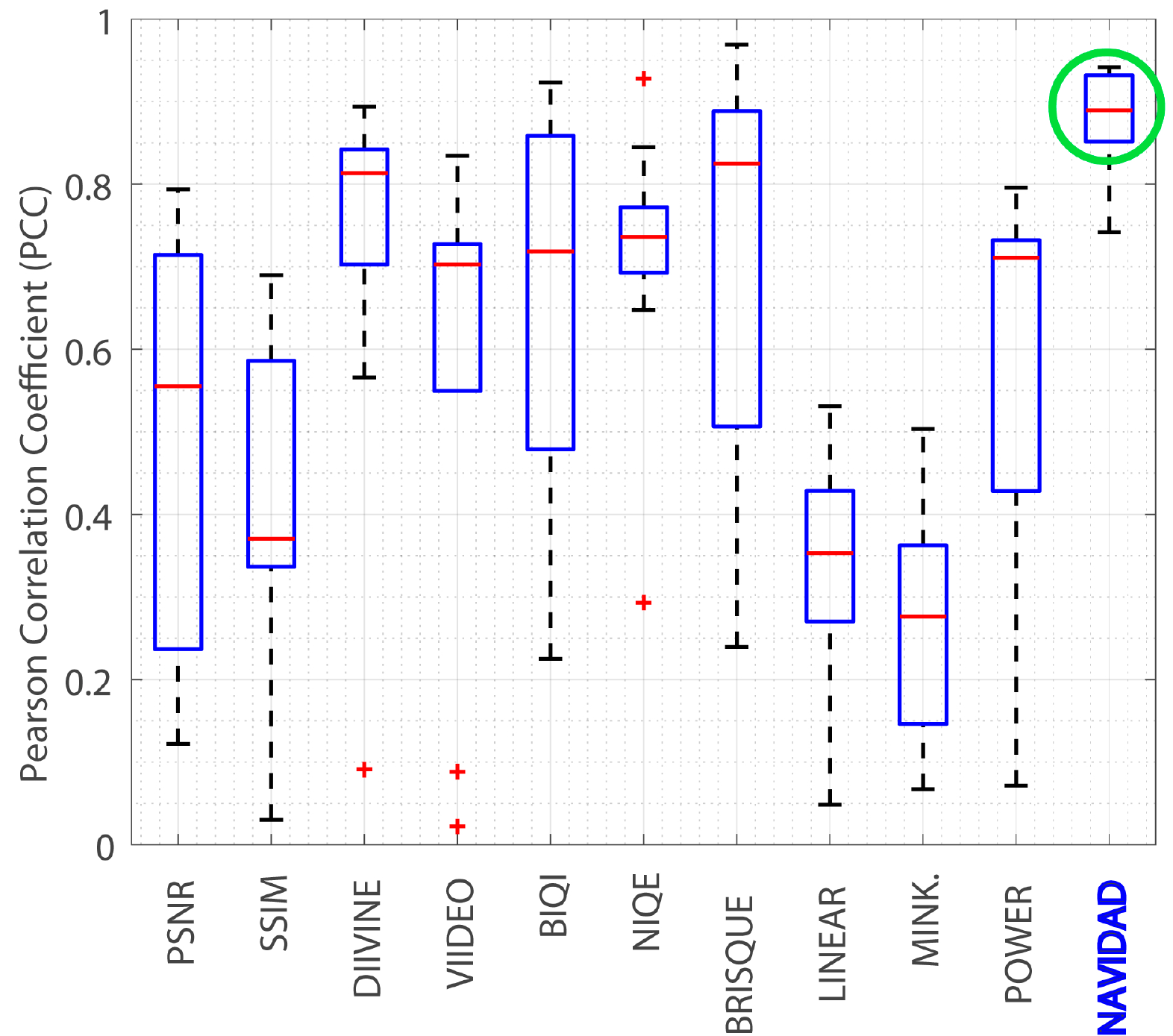}& \includegraphics[width=0.23\columnwidth]{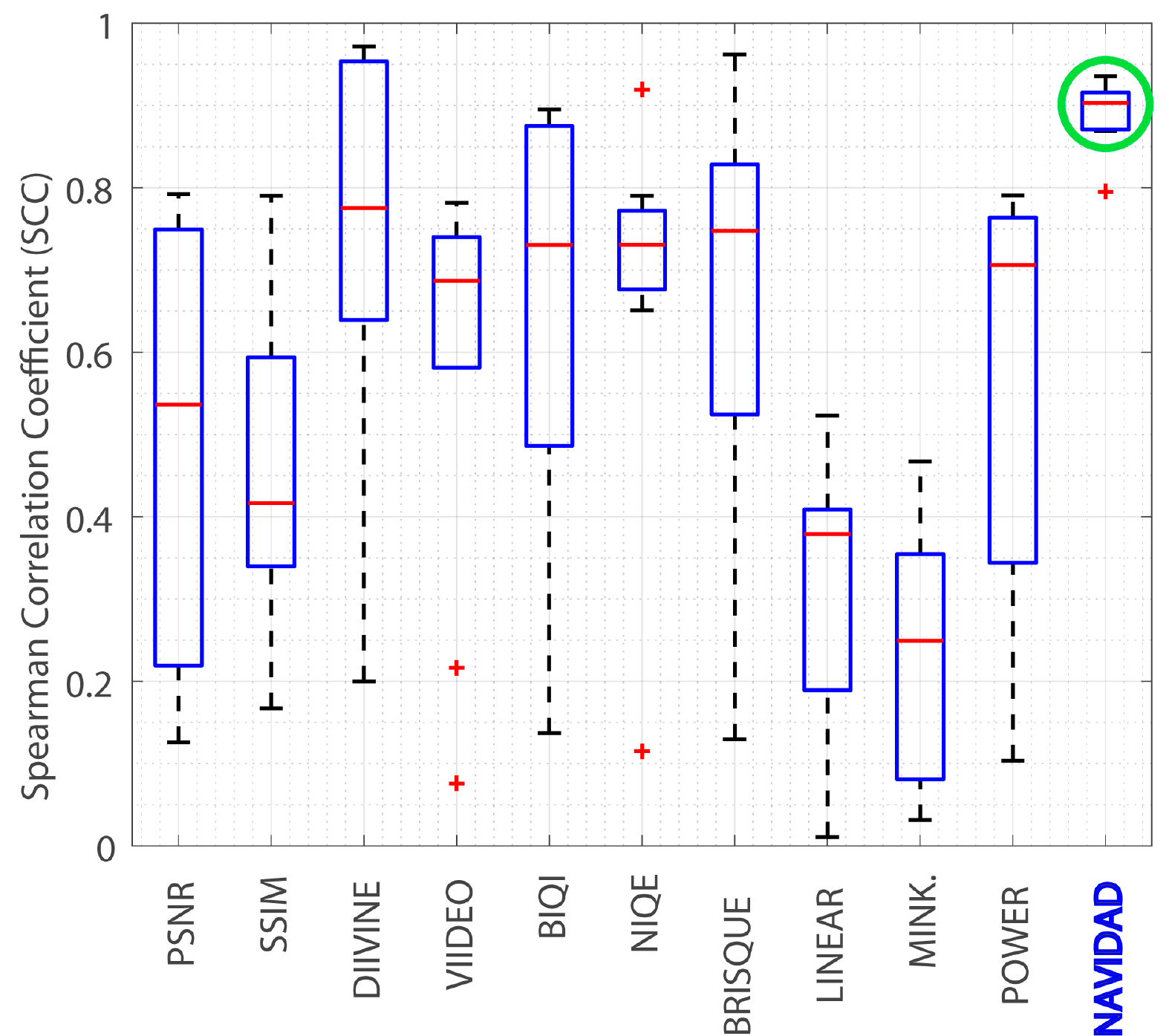}&
\includegraphics[width=0.23\columnwidth]{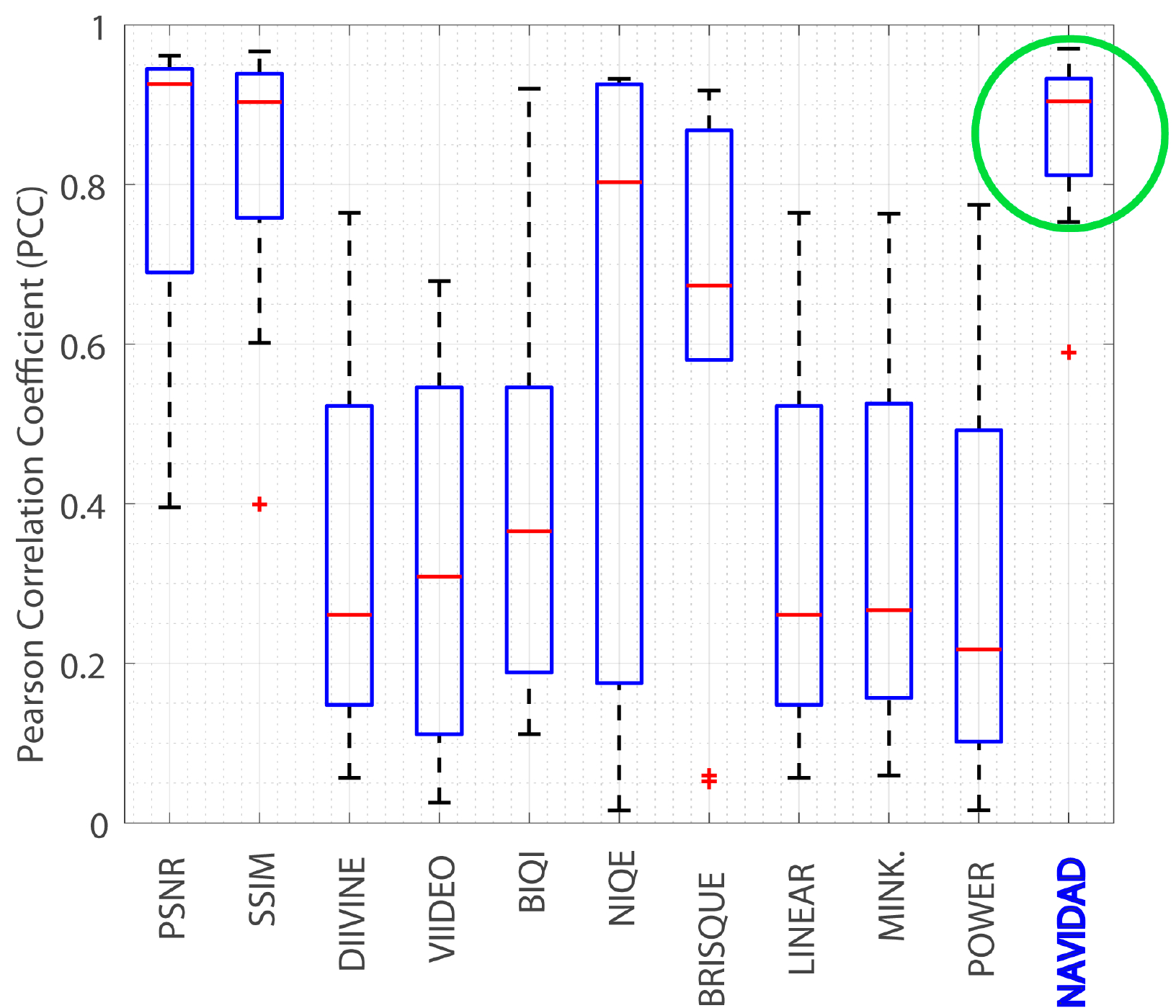}& \includegraphics[width=0.23\columnwidth]{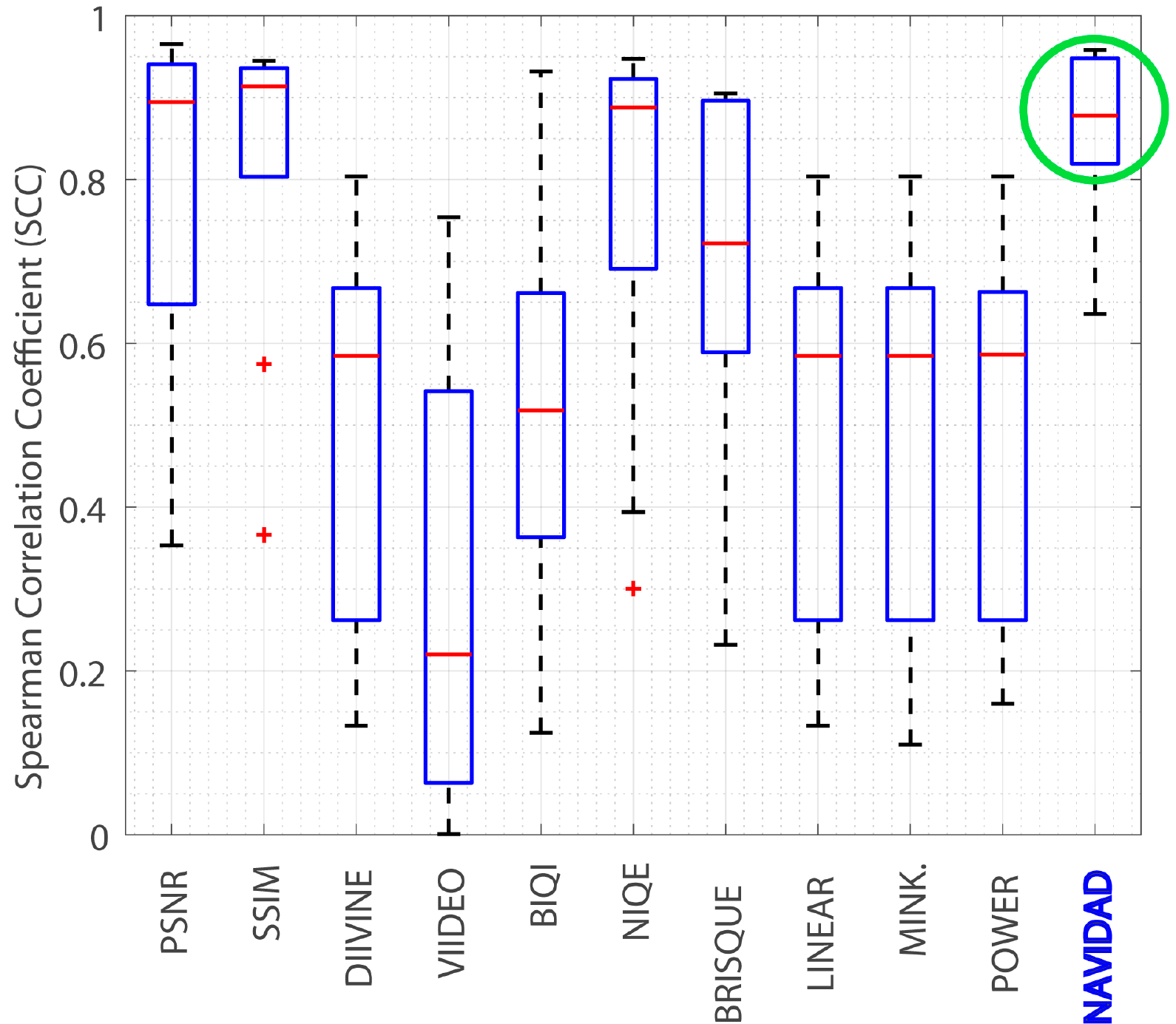}\\
\small{(a) PCC - UnB-AV}&\small{(b) SCC - UnB-AV}&\small{(c) PCC - LiveNetflix}&\small{(d) SCC - LiveNetflix}\\
\end{tabular}
\caption{Bar plot of PCC and SCC gathered from testing the litarature metrics on the UnB-AV database and the Live Netflix database.}\label{fig:res_barplots}
\end{figure*}

For a better visualization of the results, Figure \ref{fig:res_barplots} (a) and (b) depicts bar plots of the overall PCC and SCC values (over the 10 folds) for all metrics. Besides the high correlation values presented by the NAViDAd metric, it can observed that results presented a small variation on both PCC and SCC coefficients. This shows that NAViDAd's results are very consistent compared to the rest of the literature metrics.

\begin{table}[tb]
  \centering
  \caption{PCC, SCC, RMSE for the FR and NR visual and audio-visual quality metrics, tested on the UnB-AVQ Database.}
  \resizebox{.65\columnwidth}{!}{
    \begin{tabular}{rl|rr|r}
    \hline
 %\multicolumn{1}{l}{\textbf{Modality}} &
  \multicolumn{1}{l}{\textbf{Metric}} & \textbf{Measure} & \multicolumn{1}{l}{\textbf{Packet-Loss}} & \multicolumn{1}{l|}{\textbf{Frame-Freezing}} & \multicolumn{1}{l}{\textbf{All}} \\
    \hline
% \multicolumn{1}{l}{FR-Video} & 
 \multicolumn{1}{l}{PSNR} & PCC   & 0.8997 & 0.8629 & 0.7694 \\
                        & SCC   & 0.9455 & 0.8833 & 0.7368 \\
                        & RMSE  & 19.2054 & 16.5837 & 18.0728 \\ \hline
         
%           \multicolumn{1}{l}{FR-Video} &
            \multicolumn{1}{l}{SSIM \cite{wang2003multiscale}} & PCC   & 0.8563 & 0.3899 & 0.3620 \\
                       & SCC   & 0.8500 & 0.3727 & 0.3579 \\
                        & RMSE  & 2.7378 & 2.2027 & 2.4579 \\
    \hline \hline
\multicolumn{1}{l}{DIIVINE \cite{zhang2014c}} & PCC   & -0.8071 & -0.8647 & -0.8344 \\
                        & SCC   & -0.8182 & -0.5167 & -0.7519 \\
                        & RMSE  & 2.4662 & 2.9484 & 2.6939 \\ \hline
\multicolumn{1}{l}{VIIDEO \cite{mittal2016completely}} & PCC   & -0.7968 & -0.9883 & -0.8496 \\
                        & SCC   & -0.6729 & -0.9234 & -0.7834 \\
                        & RMSE  & 2.2337 & 2.6804 & 2.4449 \\ \hline
\multicolumn{1}{l}{BIQI \cite{moorthy2009modular}} & PCC   & -0.8575 & -0.9022 & -0.8310 \\
                        & SCC   & -0.9382 & -0.6000 & -0.8799 \\
                        & RMSE  & 34.8427 & 32.6918 & 33.8917 \\ \hline
\multicolumn{1}{l}{NIQE \cite{mittal2013making}} & PCC   & -0.7608 & -0.9332 & -0.8394 \\
                        & SCC   & -0.7798 & -0.7289 & -0.7195 \\
                        & RMSE  & 2.9388 & 2.4057 & 2.7119 \\ \hline
 \multicolumn{1}{l}{BRISQUE \cite{mittal2012no}} & PCC   & -0.7094 & -0.9525 & -0.8395 \\
                        & SCC   & -0.6360 & -0.9662 & -0.7728 \\
                        & RMSE  & 45.1371 & 41.4226 & 43.5049 \\ \hline
 \multicolumn{1}{l}{AV-Linear} & PCC   & 0.3919 & 0.5501 & 0.4431 \\
                       & SCC   & 0.2455 & 0.6333 & 0.3368 \\
                       & RMSE  & 10.5249 & 11.0035 & 10.7430 \\ \hline
\multicolumn{1}{l}{AV-Minkowski} & PCC   & 0.2912 & 0.4594 & 0.3422 \\
                       & SCC   & 0.2091 & 0.6333 & 0.3143 \\
                        & RMSE  & 1.9879 & 2.4289 & 2.1973 \\ \hline
\multicolumn{1}{l}{AV-Power} & PCC   & -0.6273 & -0.6938 & -0.6616 \\
                        & SCC   & -0.6727 & -0.4333 & -0.6075 \\
                        & RMSE  & 24.2614 & 23.7806 & 24.0462 \\ \hline
\multicolumn{1}{l}{{NAViDAd}} & {PCC} & \textbf{0.8638} & \textbf{0.9167} & \textbf{0.8819} \\
                        & {SCC} & \textbf{0.8773} & \textbf{0.9050} & \textbf{0.8904} \\
                       & {RMSE} & \textbf{0.5931} & \textbf{0.5718} & \textbf{0.5850} \\
    \hline
    \end{tabular}}%
      \vspace{-.5cm}
  \label{tab:av_video_results}%
\end{table}%

% Table generated by Excel2LaTeX from sheet 'AVResults'
\begin{table}[tb]
  \centering
  \caption{ PCC, SCC, and RMSE for the FR and NR audio and audio-visual quality metrics, tested on the UnB-AVQ Database.}
  \resizebox{.75\columnwidth}{!}{
    \begin{tabular}{rl|rrrr|r}
    \hline
 \multicolumn{1}{l}{\textbf{Metric}} & \textbf{Measure} & \multicolumn{1}{l}{\textbf{Noise}} & \multicolumn{1}{l}{\textbf{Chop}} & \multicolumn{1}{l}{\textbf{Clip}} & \multicolumn{1}{l|}{\textbf{Echo}} & \multicolumn{1}{l}{\textbf{All}} \\
    \hline
 \multicolumn{1}{l}{VISQOLAudio \cite{sloan2017objective}} & PCC   & 0.7945 & 0.9909 & 0.7429 & 0.6844 & 0.6008 \\
             & SCC   & 0.7000 & 1.0000 & 0.4928 & 0.5218 & 0.4781 \\
              & RMSE  & 2.4702 & 2.2047 & 2.0815 & 2.2300 & 2.2464 \\ \hline
 \multicolumn{1}{l}{VISQOL \cite{hines2015visqol}} & PCC   & 0.6102 & 0.9915 & 0.5084 & 0.4963 & 0.4236 \\
              & SCC   & 0.7000 & 1.0000 & 0.4928 & 0.5218 & 0.4645 \\
            & RMSE  & 2.6143 & 2.2045 & 2.1639 & 2.3136 & 2.3341 \\ \hline
 \multicolumn{1}{l}{PEAQ \cite{thiede2000peaq}} & PCC   & 0.7573 & 0.9347 & 0.8261 & 0.7096 & 0.7689 \\
                        & SCC   & 0.2000 & 1.0000 & 0.3189 & 0.3479 & 0.3437 \\
                        & RMSE  & 6.3196 & 5.1643 & 5.9748 & 6.0418 & 5.9704 \\
    \hline  \hline
 \multicolumn{1}{l}{P.563 \cite{Malfait2006a}} & PCC   & 0.7305 & 0.9964 & 0.9413 & 0.7752 & 0.7037 \\
                        & SCC   & 0.8000 & 1.0000 & 0.8407 & 0.4638 & 0.6367 \\
                        & RMSE  & 1.3415 & 1.3252 & 1.2310 & 1.2004 & 1.2650 \\  \hline
 \multicolumn{1}{l}{AV-Linear} & PCC   & 0.4520 & 0.9649 & 0.7718 & 0.0409 & 0.4431 \\
                       & SCC   & 0.6000 & 1.0000 & 0.3143 & -0.2571 & 0.3368 \\
                        & RMSE  & 10.9449 & 10.7825 & 10.6525 & 10.6429 & 10.7430 \\  \hline
 \multicolumn{1}{l}{AV-Minkowski} & PCC   & 0.3032 & 0.9109 & 0.6881 & -0.2842 & 0.3422 \\
                        & SCC   & 0.6000 & 1.0000 & 0.1429 & -0.2571 & 0.3143 \\
                        & RMSE  & 2.3585 & 2.2612 & 2.0770 & 2.1419 & 2.1973 \\  \hline
  \multicolumn{1}{l}{AV-Power} & PCC   & -0.7187 & -0.6990 & -0.5271 & -0.8383 & -0.6616 \\
                       & SCC   & -0.6000 & -0.5000 & -0.6000 & -0.7714 & -0.6075 \\
                        & RMSE  & 23.7961 & 24.0376 & 24.2251 & 24.0783 & 24.0462 \\  \hline
 \multicolumn{1}{l}{{NAViDAd}} & {PCC} & \textbf{0.8879} & \textbf{0.9252} & \textbf{0.8794} & \textbf{0.9044} & \textbf{0.8819} \\
& {SCC} & \textbf{0.9200} & \textbf{1.0000} & \textbf{0.8629} & \textbf{0.9086} & \textbf{0.8904} \\
& {RMSE} & \textbf{0.5764} & \textbf{0.6125} & \textbf{0.5406} & \textbf{0.6013} & \textbf{0.5850} \\
    \hline
    \end{tabular}}%
     % \vspace{-.5cm}
  \label{tab:av_audio_results}%
\end{table}%

%\subsection{Netflix Database}
To validate the proposed method, we performed a cross-validation test that consists of testing the method on an independent database, for which no training was performed. With this goal, we tested NAViDAd on the  LiveNetflix-II Database, provided by the Laboratory for Image and Video Engineering (LIVE) of the University of Texas at Austin~\cite{bampis2018towards}. This database is composed of  420  sequences, with video components at a Full HD resolution (1920$\times$1080, 4:2:0, 24 fps).  The database contains a diverse content, which includes action, documentary, video games, and sports videos.  The source videos are processed with 7 different network conditions and with 4 bitrate adaptation strategies. No audio degradations were included in the database. A total of 65 subjects rated the overall audiovisual quality of the sequences.  

Table \ref{tab:netflix-results} shows the LiveNetflix-II  results.  Since the database does not include audio degradations, we only show the results for the FR and NR visual quality metrics. It is worth pointing out that, as there are no audio degradations, the visual quality metrics have a clear advantage in this database. Unfortunately, up to our knowledge, there are no audio-visual quality databases that include both audio and video degradations.  Nevertheless,  the proposed method  performed better than the  visual (and audio-visual) quality metrics, with correlation coefficients of around 0.86. Figures \ref{fig:res_barplots} (c) and (d) present the bar plots for the average PCC and SCC values (over the 10 folds) for the tested metrics. As with the UnB-AV database, results  show that the NAViDAd correlation values varied very little across the simulations, which shows the consistency of the metric. We believe  NAViDAd can be used in  real-time streaming environments, specially in cases where audio distortions are expected to happen.  

\begin{table}[tb]
  \centering
  \caption{PCC, SCC, and RMSE gathered from testing the FR and NR visual quality metrics on the LiveNetflix-II Database.}
  \resizebox{.75\columnwidth}{!}{
    \begin{tabular}{l|rr|rrrrrr}
    \hline
          & \multicolumn{2}{c|}{\textbf{Full-Reference}} & \multicolumn{6}{c}{\textbf{No-Reference}} \\
    \hline
    \textbf{Measure} & \multicolumn{1}{l}{\textbf{PSNR}} & \multicolumn{1}{l|}{\textbf{SSIM}} & \multicolumn{1}{l}{\textbf{DIIVINE}} & \multicolumn{1}{l}{\textbf{VIIDEO}} & \multicolumn{1}{l}{\textbf{BIQI}} & \multicolumn{1}{l}{\textbf{NIQE}} & \multicolumn{1}{l}{\textbf{BRISQUE}} & \multicolumn{1}{l}{\textbf{NAViDAd}} \\
    \hline
    \textbf{PCC} & 0.6981 & 0.7333 & -0.8364 & -0.6598 & -0.4263 & -0.7550 & -0.7271 & \textbf{0.8611} \\
    \textbf{SCC} & 0.6911 & 0.7123 & -0.8106 & -0.7153 & -0.4724 & -0.7701 & -0.7115 & \textbf{0.8599} \\
    \textbf{RMSE} & 32.2445 & 2.3024 & 2.6126 & 2.5265 & 38.3084 & 3.8324 & 56.2907 & \textbf{0.5929} \\
    \hline
    \end{tabular}}%
    \vspace{-.2cm}
  \label{tab:netflix-results}%
\end{table}%

\section{Conclusions}\label{sec:conclusion}
In this work, we proposed a no-reference audio-visual quality metric, which is base on a deep autoencoder technique. The proposed model used a set of audio and video features to estimate the overall audiovisual quality. The model is formed by a 2-layer framework that includes  a deep autoencoder layer and a classification layer. These two layers are stacked and trained  to build the deep neural network model. These feature sets were passed  to a two-layer autoencoder that produced a set of features with lower dimension. Then, a classification function mapped these features into subjective scores. %As a final stage, the output of the model was processed to compute the overall audiovisual quality. 
Results showed that the proposed approach estimates the perceived quality of audio-visual sequences with good accuracy. The model presented a significant advantage when compared to several video, audio, and audio-visual objective metrics from the literature. The model was also tested on a external database and also presented a good performance. We believe better results can be achieved  with the refinement of the network parameters. {In addition, further tests (e.g. an ablation study) can be made in order to determine the relative importance of audio and video features in the proposed system.}
%Results showed that the audiovisual model and the deep autoencoder approach, presented in this work, are able to predict the perceived quality of audio-visual sequences at a good correlation level when compared to several video and audio objective metrics from the literature.

%\section*{Acknowledgments}

\bibliographystyle{IEEEtran}
\bibliography{bibliografia}

\end{document}